\documentclass{elsart3p}
\usepackage{amssymb}
\usepackage{epsfig}

\begin{document}

\begin{frontmatter}

\title{Superconducting phase formation in random neck syntheses: a study of the Y-Ba-Cu-O system by magneto-optics and magnetometry }

\author[UNIBE]{J.B. Willems}
\author[MPI]{, J. Albrecht}
\author[UNIBE]{, I.L. Landau} 
\author[UNIBE]{, J. Hulliger}

\address[UNIBE]{Department of Chemistry and Biochemistry, University of Berne, Freiestrasse 3, CH-3012-Berne, Switzerland}
\address[MPI]{Max-Planck-Institute for Metals Research, Heisenbergstr. 3, D-70569 Stuttgart, Germany}

\begin{abstract}

Magneto-optical imaging and magnetization measurements were applied to investigate local formation of superconducting phase effected by a random neck synthesis in Y-Ba-Cu-O system. Polished pellets of strongly inhomogeneous ceramic samples show clearly the appearance of superconducting material in the intergrain zones of binary primary particles reacted under different conditions. Susceptibility measurements allows evaluation of superconducting critical temperature, which turned out to be close to that of optimally doped YBa$_2$Cu$_3$O$_{7-x}$.

\end{abstract}

\begin{keyword}
high-$T_c$ superconductivity \sep YBa$_2$Cu$_3$O$_{7-x}$ \sep magneto-optical imaging \sep random neck syntesis \sep solid state chemistry
% PACS codes here, in the form: \PACS code \sep code
\PACS 74.60.-w \sep 74.-72.-h
\end{keyword}
\end{frontmatter}

\section{Intrduction}

Recently, a combinatorial approach  for searching new oxide superconductors was presented \cite{1,2,hats,4}. The basic idea is to generate most inhomogenously reacted ceramic samples in order to synthesize new superconducting compounds \cite{1,2,hats}. Starting with a mixture of grains of several initial components, due to their statistical distribution, various combinations of original grains are created in different parts of a sample. If such a mixture is heated up to temperatures sufficient to start a chemical reaction between neighboring grains,  one can obtain a significant variety of resulting products. After reaction, the resulting ceramic material can be ground to small particles and a magnetic separation procedure may be used in order to selectively take out grains exhibiting superconductivity \cite{4,5}. A key issue for the functioning of such an approach is to find the conditions, which would lead to the formation of superconducting phases within volumes large enough for further processing. In this work, we employ magneto-optical imaging (MOI) as a technique to visualize superconducting regions in inhomogeneously reacted samples. We demonstrate that by a random neck synthesis (RNS) \cite{5} in a mixture of two component ($i/j$) prereacted BaO$_2$/Y$_2$O$_3$, BaO$_2$/CuO and CuO/Y$_2$O$_3$ particles, superconducting YBa$_2$Cu$_3$O$_{7-x}$ with $T_c \approx 92$ K can be obtained and detected. Superconducting regions appear embedded in a matrix of non-supercoducting products of the reaction.  

\section{Experimental}

The magneto-optical (MO) Faraday effect allows visualization of the magnetic field distribution down to micrometer resolution \cite{6}. Due to expulsion of the field from superconducting regions, they can be identified in MO images  of non-uniform samples. A film made of lutetium doped ferrimagnetic iron garnet \cite{7} was used as a MO layer. In order to visualize the magnetic field distribution, the MO layer was placed directly onto the polished surface of a ceramic sample. Linearly polarized light for illumination and a microscope equipped with a CCD camera were used for observation and data acquisition. An additional polarizer situated between the microscope and the sample served to obtain a desired MO contrast. More details about the experimental set-up can be found elsewhere \cite{8}. Previous studies showed that MOI can be rather effective for an investigation of superconducting properties of polycrystalline materials (see, e.g., Ref. \cite{9}).

Magnetization measurements were carried out on a SQUID Magnetometer with a 5 T magnet
(Quantum Design).

\section{Sample preparation}

In view of a ceramic combinatorial approach providing conditions for maximum chemical diversity within a single sample \cite{1,2,hats,4}, three binary ($i/j$)  combinations of Cu/Y, Cu/Ba and Ba/Y oxides were prereacted by classical solid state chemistry. First, CuO, Y$_2$O$_3$ and BaO$_2$ were mixed according to compositions BaO$_2$$\cdot$CuO, 2CuO$\cdot$Y$_2$O$_3$ and 2BaO$_2$$\cdot$Y$_2$O$_3$. A homogeneous distribution of components was achieved by joint ball milling of the corresponding ($i/j$) oxide mixtures. The resulting powders were pressed into disk shaped pellets. These pellets were heated in air at 950 $^{\circ}$C for 24 h. 2BaO/Y$_2$O$_3$ pellets were then heated for additional 24 h at 1100 $^{\circ}$C.

Each set of pellets belonging to one of three element combinations were independently transformed into particles of about 80 - 100 $\mu$m. This was done by breaking the pellets in a mortar and sieving out the corresponding size fraction. In next step, equal amounts of these three types of ($i/j$) particles were thoroughly mixed together and this mixture was used for making samples, i.e.,  pressed into pellets and heated at 930 $^\circ$C in an oxygen atmosphere. Heating and cooling rates were 300 $^\circ$C/h and 15 $^\circ$C/h, respectively. The sample S1 was kept at 930 $^\circ$C for one week and the sample S2 for three weeks. 

%%%%%%%%%%
\begin{figure}[h]
 \begin{center}
  \epsfxsize=0.8\columnwidth \epsfbox {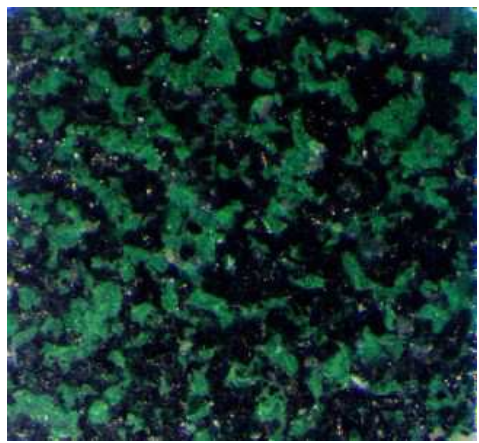}
  \caption{A photographic image of the sample S2 after polishing (horizontal size corresponds to 3 mm). Grey areas correspond to the green phase (Y$_2$BaCuO$_5$).}
   \label{sample_im}
 \end{center}
\end{figure}
%%%%%%%%%%
After such a procedure pellets were fragile. In order to improve their stability and make them suitable for MOI experiments, they were placed in an epoxy resin EPO-TEK 305 and polymerized  under a N$_2$ pressure of 6 bar. After removal of excess resin, the flat surfaces were polished with a fine diamond powder (1 $\mu$m particle size). This was necessary in order to obtain a close contact between the sample surface and the MO layer, which is essential to ensure a high spatial resolution. A photograph of the sample S2 is shown in Fig. \ref{sample_im}. It may be noted that both samples looked almost identical. 

The basic idea behind the RNS approach is to provide reactive contacts between grains of different compositions. If volumes of starting grains are not used up during the reaction time, all phases, which may exist in the Y-Ba-Cu-O system at given conditions (temperature, total pressure and partial pressure of oxygen) should be formed in neck zones between different grains for thermodynamic reasons. As original particles were chosen sufficiently large, product particles may also  acquire a size suitable for further handling and analyses. 

\section{Results}

\subsection{MO imaging}

%%%%%%%%%%
\begin{figure}[h]
 \begin{center}
  \epsfxsize=0.8\columnwidth \epsfbox {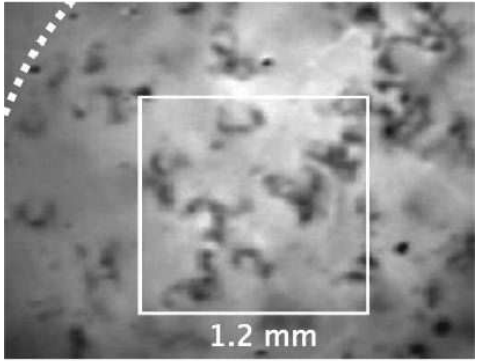}
  \caption{MO image of the sample S1 obtained at $T = 10$ K and $H = 160$ Oe after zero-field cooling. Superconducting areas are visible as dark structures. The dashed line corresponds to an edge of the sample. The white square indicates a part of the image, which is shown in Fig. 3. }
   \label{MOimage}
 \end{center}
\end{figure}
%%%%%%%%%%
One of the MO images of the sample S1 is shown in Fig. \ref{MOimage}. The sample was cooled  down to $T = 10$ K in a zero magnetic field. Then, the magnetic field $H = 160$ Oe was applied perpendicularly to the sample surface. Crossed polarizer and analyzer were used for observation. In this case, areas with a non-zero magnetic field are seen as {\it bright}, while superconducting regions, due to expulsion of magnetic flux, can be identified as {\it darker} spots. As expected, superconducting inclusions are distributed irregularly throughout the sample. 

%%%%%%%%%%
\begin{figure}[h]
 \begin{center}
  \epsfxsize=1\columnwidth \epsfbox {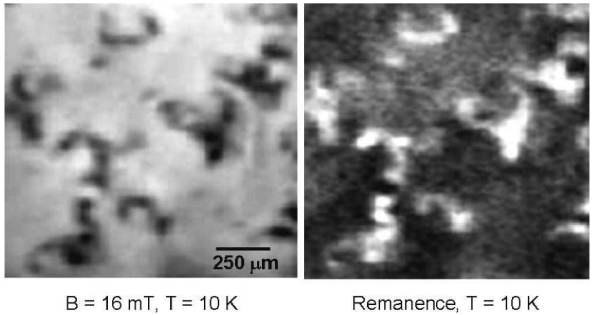}
  \caption{(left panel) Enlarged part of the image shown in Fig. \ref{MOimage} (zero-field cooling). (right panel) The same area of the sample in the remanent state after rising $H$ to 2.5 kOe and then reducing to zero.}
   \label{remanent}
 \end{center}
\end{figure}
%%%%%%%%%%
An area, indicated in Fig. \ref{MOimage} by the white square, is shown in the left panel of Fig. \ref{remanent} at higher magnification. After taking this image, the field was increased to 2.5 kOe and then switched off. The resulting image of the same area is shown in the right panel of Fig. \ref{remanent}. As may be seen, dark spots became bright after switching off the magnetic field. This is clear evidence of magnetic flux trapped inside these areas, which is a secure identification of superconductivity.

%%%%%%%%%%
\begin{figure}[h]
 \begin{center}
  \epsfxsize=1\columnwidth \epsfbox {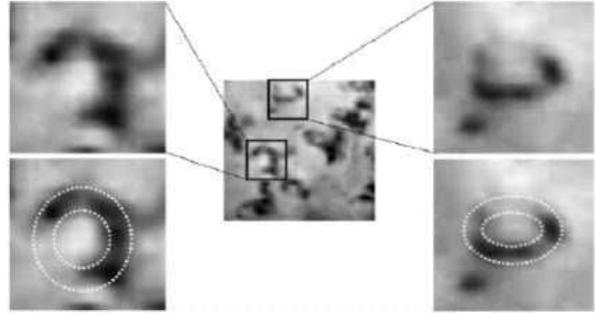}
  \caption{Magnified areas of Fig. \ref{remanent}. Superconducting regions exhibit typical shapes of intergrain contacts with the superconducting phase formed around some particular grains as it is indicated by the dotted lines. }
   \label{MOdetails}
 \end{center}
\end{figure}
%%%%%%%%%%
Having a closer look on the flux density distribution shown in Figs. \ref{remanent} and \ref{MOdetails}, some particular features are apparent. The most extended superconducting regions exhibit lateral dimensions of several hundred micrometers, being substantially larger than typical grains that were preliminary chosen (80-100 $\mu$m). Another interesting feature is that the shape of superconducting regions provides convincing evidence that the superconducting phase was predominantly formed around some sufficiently big grains. Fig. \ref{MOdetails} underlines this fact. We also note that dark areas of the image, corresponding to the superconducting regions, are not uniform but consist of smaller black spots connected by dark grey areas. This reflects non-uniform superconductivity inside superconducting regions. One can imagine that, during the reaction, a superconducting phase was initially formed in several closely situated  reaction centers and their subsequent extension resulted in the formation of clusters consisting of several superconducting grains with somewhat weaker superconducting links between them. As we discuss below, similar conclusions can be drawn from the analysis of magnetization data.

%%%%%%%%%%
\begin{figure}[!t]
 \begin{center}
  \epsfxsize=1\columnwidth \epsfbox {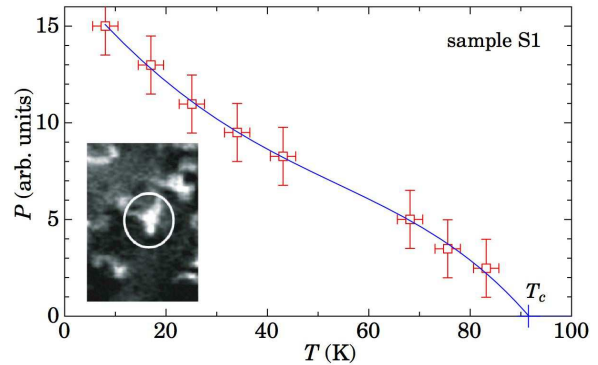}
  \caption{ The MO contrast $P$ as a function of temperature. The results were obtained in the remanent state as shown in Fig. \ref{remanent}. The solid line is a guide to the eye drawn in a way that $P$ vanishes at $T_c = 91.6$ K evaluated from magnetization measurements. The inset shows a part of the MO image, which was used for evaluation of $P$. }
  \label{contrast}
 \end{center}
\end{figure}
%%%%%%%%%
In order to quantify the results of MOI experiments, we introduce the magneto-optical contrast $P$ as the the difference between light intensities coming from superconducting and normal regions of the sample. The particular superconducting region, which was used for calculation of $P(T)$ data, is indicated by the white circle in the inset of Fig. \ref{contrast}. The remanent state was created at $T = 10$ K as described above. Then, $P$ was measured as a function of increasing temperature. The results are plotted in Fig.  \ref{contrast}. Obviously, $P$ must vanish at $T = T_c$. As may be seen, results of MOI experiments are in good agreement with $T_c = 91.6$ K, evaluated from magnetization data. 

%%%%%%%%%%
\begin{figure}[!h]
 \begin{center}
  \epsfxsize=0.7\columnwidth \epsfbox {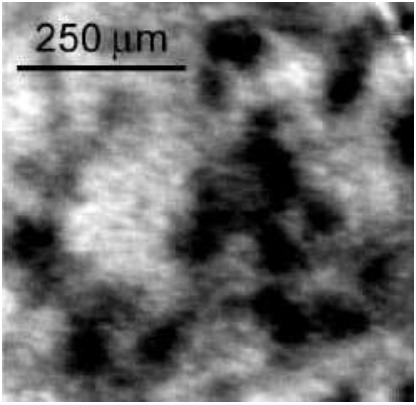}
  \caption{ A MO image of the sample S2 at $T = 10$ K and $H = 40$ Oe after zero-field cooling. Superconducting regions are dark.}
  \label{sample2}
 \end{center}
\end{figure}
%%%%%%%%%
A MO image of the sample S2, which underwent a much longer thermal treatment,  is shown in Fig. \ref{sample2}. Although this image is similar to those of the sample S1, there are considerable differences. Dark spots corresponding to superconducting regions have simpler shapes than those shown in Fig. \ref{remanent}. In addition to almost black superconducting grains, there are extended dark grey areas. Most likely that these regions correspond to similar superconducting grains, which are situated at some distances below the sample surface and for this reason they provide a weaker MO contrast.  

\subsection{Magnetization measurements}

 %%%%%%%%%%
\begin{figure}[h]
 \begin{center}
  \epsfxsize=1\columnwidth \epsfbox {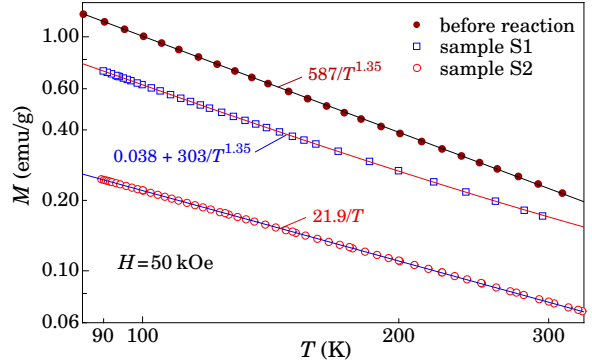}
  \caption{ $M(T)$ curves measured in $H = 50$ kOe. The solid lines are the best linear fits with $M(T) = m_0 + m_1/T^n$ with three adjustable parameters $m_0$, $m_1$ and $n$. The resulting values of fit parameters are indicated near the curves.  }
  \label{normal}
 \end{center}
\end{figure}
%%%%%%%%%
Magnetic properties of these multicomponent samples were rather complex. The normal-state $M(T)$ curves measured in $H = 50$ kOe are plotted in Fig. \ref{normal} on log-log scales. The curve corresponding to the non-reacted mixture of components is also shown. As may be seen, the $M(T)$ dependences for the original mixture and for the sample S2 can very well be described by a simple power law $1/T^n$. While the sample S2 follows the Curie-Weiss law ($n = 1$), the corresponding power for the original mixture $n \approx 4/3$. The $M(T)$ curve for the sample S1 is not a straight line on these scales. In order to approximate $M(T)$ for this sample, we used $M(T) = m_0 + m_1/T^n$. Fit to experimental data points results in the same value of $n \approx 4/3$ as for the original mixture (see Fig. 7).

%%%%%%%%%%
\begin{figure}[h]
 \begin{center} 
  \epsfxsize=1\columnwidth \epsfbox {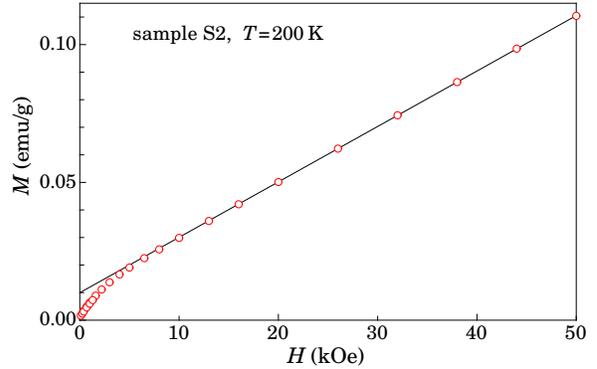}
  \caption{ Magnetization as a function of magnetic field at $T = 200$ K. The solid line is the best linear fit to $M(H)$ data points for $H \ge 13$ kOe. }
  \label{S2_norm}
 \end{center}
\end{figure}
%%%%%%%%%
The magnetic field dependence of the normal-state magnetization for the sample S1, as well as for the original mixture, were purely paramagnetic with $M(H) = \chi_n H$, where $\chi_n$ is the normal-state magnetic susceptibility. The sample S2 was different. As may be seen in Fig. \ref{S2_norm}, this sample had  a noticeable ferromagnetic component. The high field part of the $M(H)$ curve could be described by $M(H) = M_0 + \chi_n H$. The value of $M_0$ was practically independent of temperature.

%%%%%%%%%%
\begin{figure}[h]
 \begin{center}
  \epsfxsize=1\columnwidth \epsfbox {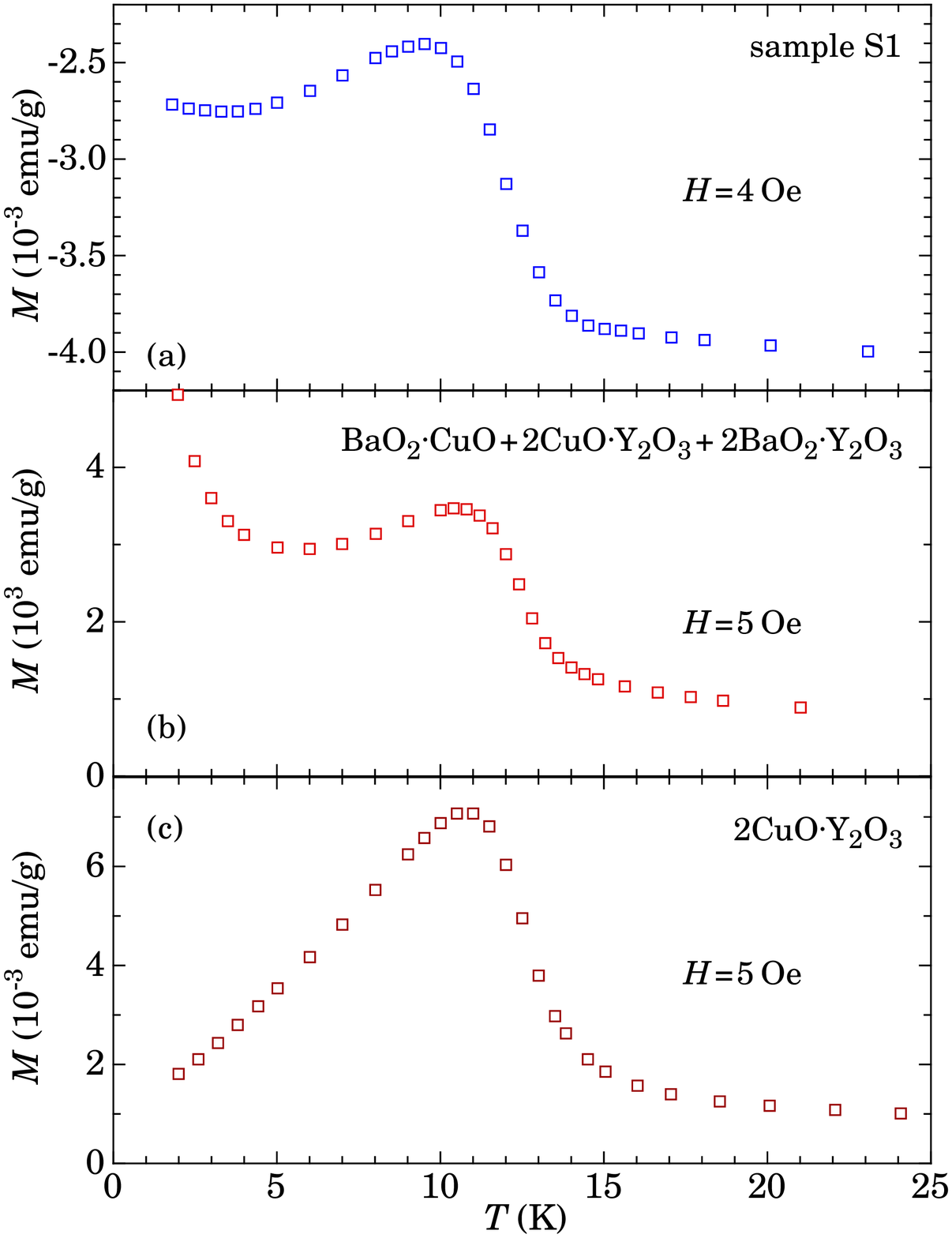}
  \caption{ The $M(T)$ curves measured in low applied fields. (a) Sample S1. (b) A non-superconducting mixture of oxides before thermal treatment. (c) A prereacted 2BaO$\cdot$Y$_2$O$_3$ ($i/j$) component.}
  \label{transition}
 \end{center}
\end{figure}
%%%%%%%%%
A low-temperature part of the $M(T)$ curve for the sample S1 is shown in Fig. \ref{transition}(a). At these low temperatures, the mixed-state contribution to $M$ is temperature independent and all temperature variations of $M$ are due to non-superconducting phases of the sample. There is a clear magnetic (antiferromagnetic) transition at temperatures a little bit above 10 K (Fig. \ref{transition}(a)). This transition was apparently originated from Y$_2$Cu$_2$O$_5$ (blue phase), which was present in a 2BaO$\cdot$Y$_2$O$_3$ component (Fig. \ref{transition}(c)) \cite{YCuO}. The transition could also be observed in the non-superconducting mixture of oxides before a thermal treatment (Fig. \ref{transition}(b)). Contrary to that, the sample S2, which underwent 3 weeks of thermal treatment, did not show any traces of this transition. 

%%%%%%%%%%
\begin{figure}[h]
 \begin{center}
  \epsfxsize=1\columnwidth \epsfbox {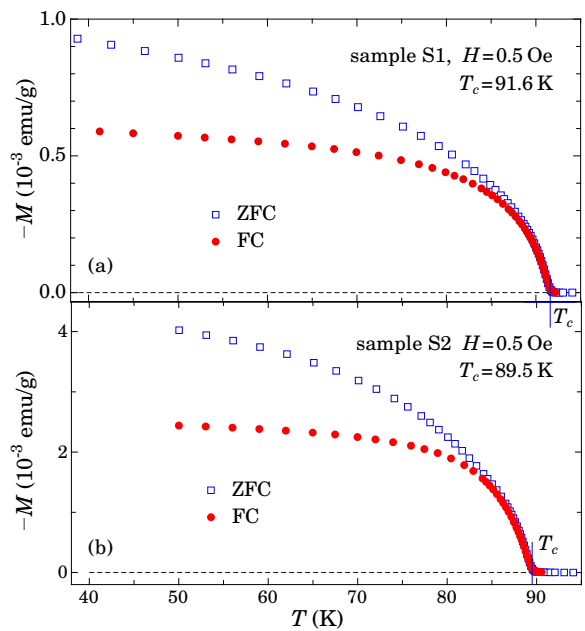}
  \caption{ZFC and FC magnetization curves.  (a) Sample S1. (b) Sample  S2.}
  \label{S1_magn}
 \end{center}
\end{figure}
%%%%%%%%%
Zero-field cooled (ZFC) and field-cooled (FC) magnetization curves for the samples S1 and S2 measured in $H= 0.5$ Oe are shown in Fig \ref{S1_magn}(a) \ref{S1_magn}(b). The results for both samples are similar. Magnetization is practically reversible at temperatures very close to $T_c$ with an onset of irreversibility at lower temperatures. The critical temperature can be evaluated as a linear extrapolations of the steepest part of the $M(T)$ curve to $M = 0$. This results in $T_c = 91.6$ K and $T_c = 89.5$ K for the samples S1 and S2, respectively. 

\begin{figure}[h]
 \begin{center}
  \epsfxsize=1\columnwidth \epsfbox {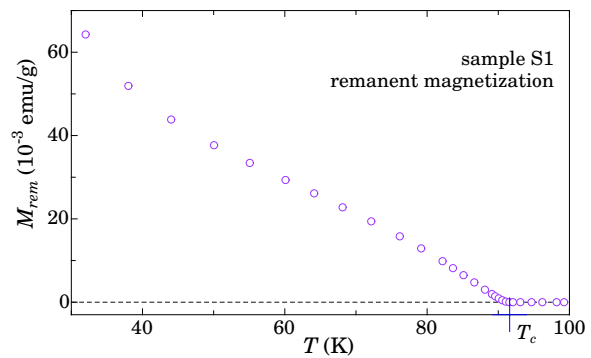}
  \caption{ Remanent magnetization as a function of temperature. The remanent state was created at $T = 10$ K by switching on a field $H = 2.5$ kOe and switching it off, as it was done for MOI study (see Figs. \ref{remanent} and \ref{contrast})}
  \label{remmagnn}
 \end{center}
\end{figure}

Fig. \ref{remmagnn} shows the temperature dependence of the remanent magnetization ($M_{rem}$) measured for the same conditions as the $P(T)$ curve presented in Fig. \ref{contrast}. $M_{rem}$ vanishes at $T = 91.4$ K, just a bit below $T_c = 91.6$ K. Comparing $M_{rem}(T)$ and $P(T)$ curves, one should take into account that the MO contrast is proportional to the magnetic induction $B$ in the center of a superconducting region, while $M_{rem}$ is proportional to the integral of $B$ over the sample volume. This difference between  $M_{rem}$ and $P$ results in slightly different shapes of the $M_{rem}(T)$ and $P(T)$ curves, as may be seen in Figs. \ref{contrast} and \ref{remmagnn} (see also \cite{adb}). 
%%%%%%%%%%
%%%%%%%%%

\section{Discussion} 

MOI provides direct information on characteristic sizes and shapes of superconducting regions. As may be seen in Figs. 3 and 6, while the sizes of superconducting regions are approximately the same for both samples, their shapes are quite different.  Also the total amount of superconducting phase is considerably higher for the sample S2.  

Magnetization measurements provide some complementary information. For instance, the evolution of the magnetic transition (see Fig. 9) and its disappearance after three weeks of the reaction time allows to conclude that initial 2BaO$\cdot$Y$_2$O$_3$ grains, which were still preserved after 1 week of thermal treatment, were completely used up after 3 weeks. We also note a ferromagnetic contribution to magnetization (see Fig. 8), which was observed in the sample S2 and was absent in S1. This is an unexpected result, which shows that some new phases can be created only after many days of thermal treatment.

Considering the magnetization curves presented in Fig. 10, we  point out magnetic reversibility at temperatures very close to $T_c$. This reversibility disappears if the magnetic field is increased to 3-5 Oe. Because pinning effects are especially strong in low magnetic fields, this behavior cannot be attributed to the vortex motion. The only plausible explanation is that superconducting grains are too small to capture even a single vortex line in a field cooling process. As it is discussed in more detail in \cite{lwh}, in order to create the mixed state in a superconducting grain, its size $d$ must be at least twice larger than $D_0 = \sqrt{\Phi_0/H}$ ($\Phi_0$ is the magnetic flux quantum). Such an estimate gives $d \sim 10$ $\mu$m. This value of $d$ is, however, substantially smaller than it can be evaluated from MO images presented in Figs. 3 and 6. We believe that $T_c$ is not uniform across superconducting regions. This is why superconducting domains may be rather small at higher temperatures. With decreasing temperature, superconducting regions grow in size and start to overlap creating superconducting clusters, which can capture magnetic flux and lead to irreversibility. 

While the critical temperature for the sample S1 is close to an optimally doped YBCO (see Fig. 10(a)), $T_c$ for the sample S2 is about 2 K lower. Because during the reaction time the samples were held in an oxygen atmosphere, we consider it unlikely that the samples can be oxygen deficient . It is more probable that non-stoichiometry of other elements is a true reason for this effect.

\section{Conclusions}

The present analysis demonstrates the analytical ability of MOI for the investigation of local magnetic properties of inhomogeneous ceramic samples. Combination of MOI with magnetization measurements provides even deeper insight into internal structure of superconducting regions and their evolution with reaction time.  Application of the RNS principle to a well known system allowed to demonstrate that local formation of superconducting material is possible and that the use of two component ($i/j$) starting grains with sizes $\sim 80-100$ $\mu$m can yield product volumes which are suitable for magnetic separation and structural characterization. Recently, we have presented optical results on local product formation when using micrometer sized starting materials \cite{cwt}. Present results provide a realistic base for the application of RNS in combination with magnetic separation [4,5] for the search for minority phases of superconducting materials featuring higher $T_c$ values than known for the bulk of a ceramic sample.

\ack

This work was supported by the Swiss NCCR MaNEP II under project 4, novel materials.

\end{document}